\documentclass[12pt, fleqn]{article}
\usepackage[cp1251]{inputenc}
\usepackage{latexsym,amsfonts,amssymb}

\newtheorem{theo}{Theorem}
\newcommand{\bt}{\begin{theo}}
\newcommand{\et}{\end{theo}}
\newcommand{\bd}{\begin{displaymath}}
\newcommand{\ed}{\end{displaymath}}

\newcommand{\lf}{\left}
\newcommand{\rg}{\right}

\newcommand{\be} {\begin{equation}}
\newcommand{\ee} {\end{equation}}
\newcommand{\ba} {\begin{array}}
\newcommand{\ea} {\end{array}}

\newcommand{\lbd} {\lambda}

\begin{document}

 {\Large \bf New conditional symmetries \\
 and exact solutions of
 nonlinear \\
 reaction-diffusion-convection equations. I
 }\\
 \medskip\\
{\bf Roman Cherniha}\footnote{\small e-mail: cherniha@imath.kiev.ua
} {\bf and  Olexii Pliukhin}\footnote{\small e-mail:
pliukhin@imath.kiev.ua } \\
{\it  Institute of Mathematics, Ukrainian National Academy
of Sciences\\
 Tereshchenkivs'ka Street 3, Kyiv 01601, Ukraine}

\renewcommand{\abstractname}{}
\begin{abstract}
A complete description of $Q$-conditional symmetries  for two
classes of reaction-diffusion-convection equations with power
diffusivities is derived. It is shown that  all the known    results
for reaction-diffusion equations with power diffusivities follow as
particular cases from those obtained here but not vise versa.

\end{abstract}

\begin{center}
{\bf 1.Introduction.}
\end{center}

Nonlinear reaction-diffusion-convection (RDC) equations of the form
\be\label{*}
 U_t=\left[A(U)U_x\right]_x+ B(U)U_x+ C(U),
\ee where $U= U(t,x)$ is the unknown function, $  A(U), B(U), C(U)$
are  given  smooth functions and the subscripts $t$ and $x$ denote
differentiation with respect to these variables, arise in a wide
range of mathematical models describing various processes in physics
and  biology  \cite{ames,mur, mur2}. Starting from the remarkable
Ovsiannikov work \cite{ovs-1959}  a great number of papers devoted
to investigation of these equations by means of group-theoretical
methods. At the present time one can claim that all Lie symmetries
of   (\ref{*}) are completely described and the relevant Lie
solutions are constructed for equations of the form
 (\ref{*}), which arise in applications (see \cite{dor}--\cite{ch-se-2006} and the papers cited therein).

In 1969 Bluman and Cole \cite{bl-c} introduced an essential
generalization of Lie symmetry using the simplest  representative of
(\ref{*}), the linear heat equation. These generalized  symmetries
are often  called nonclassical symmetries  nevertheless this notion
was not  used in  \cite{bl-c}. The  notion  of  nonclassical
symmetry was further developed in \cite{Olve87} and \cite{Levi89}. A
new generalization of Lie symmetry, conditional symmetry, was
suggested by Fushchych and his collaborators \cite{Fush88}, \cite
[Section 5.7]{Fush93}. Note that notion nonclassical symmetry can be
derived as a particular case from conditional symmetry but not vise
versa (see, e.g., an example in \cite{ch-he-2004}). In the middle of
90-s of the last century the notion of generalized conditional
symmetry was  introduced  \cite{Foka94},
 which again can be considered as a special case of
conditional symmetry. Taking this  into account, to avoid any
misunderstanding we continuously use the terminology $Q$-conditional
symmetry  instead of nonclassical symmetry. In fact, there are
several types non-Lie symmetries at the present time and each of
them can be called
 nonclassical one.

While there is no existing general theory for integrating nonlinear
RDC of the form (\ref{*}),
 construction of  particular exact solutions for these equations
 is  a non-trivial and  important problem. Finding  exact solutions that have a
physical, chemical or biological
 interpretation is of fundamental importance.
It is well-known that  the notion of $Q$-conditional symmetry plays
an important role in investigation of nonlinear RDC equations since,
having such symmetries in the explicit form, one may construct new
exact solutions, which are not  obtainable  by the classical Lie
machinery.   Several papers were devoted to this topic during the
last 15 years \cite{ch-se-98, Fush93, se-90,  nucci1, cla, a-h,
che-2006}.
 The time is therefore ripe for a complete description
 of non-Lie symmetries for the general  RDC equation  (\ref{*}).
Since it  seems to be extremely difficult task at the present time
here  we present the solving
 for some important particular cases of (\ref{*}), namely:
 \be \label {1}
 {U_t}= \ [U ^{m} U_{x}]_{x}+  \lambda U^{m}
 U_{x}+C(U),\ee
 \be \label{1b} U_t=[U^mU_x]_x+\lbd
U^{m+1}U_x+C(U),\ee
 where
 $\lambda$ and $ m$ are arbitrary constants while  $C(U)$ is an arbitrary functions.

 It should be noted that $Q$-conditional symmetry of (\ref{1}) with $\lambda=0,\ m=0$,
 i.e. reaction-diffusion equation, was investigated in \cite{Fush93, se-90}, the most general results were obtained
 in \cite{ cla, dix-cla}. Operators of $Q$-conditional symmetry  for  equation (\ref{1})
  with $\lambda=0,\ m\ne0$ have been constructed in   \cite{a-h} while
    the  complete description of
  the RDC  (\ref{1b}) with $ m=0$  is presented  in the recently published paper  \cite{che-2006}. Finally, we remind the reader that the
  determining equations for constructing the $Q$-conditional symmetry operators  of
  the general  RDC equation  (\ref{*}) were obtained in \cite{ch-se-98}, however, that paper contains only examples
  of particular solutions of those equations.

 We stress that only reaction-diffusion equations with the convective  terms ($\lambda\neq0$)
 are considered below. The motivation of this restriction  has two aspects.  The first one
  is  to find  $Q$-conditional symmetries for  nonlinear   equations   involving
  three transport mechanisms (diffusion, reaction and convection) in contrary to standard reaction-diffusion
   (RD) equations.
  The second one is to deal with the equations, which arise in applications.  In fact, sometimes the convection arises as a natural
extension of a conservation low and then
  one obtains RDC equations instead of  RD equations. The effect of nonlinear convection in RD equations can have
  "a dramatic effect on solutions" \cite[Section 11.4]{mur}).

 The paper is organized as follows.
  In the second section, we present two theorems  giving
a complete description of $Q$-conditional symmetries of the
nonlinear RDC equations (\ref{1})-(\ref{1b}).
  In the third section, the proof of the theorems are presented.
  The main results of the
paper are summarized and discussed in the last section. In the
second part of this paper, we shall apply
   the $Q$-conditional symmetries obtained  for constructing  new exact
solutions of the RDC equations arising in various applications.

\begin{center}
{\textbf{2. Main results.}} \end{center}

We want to find all possible $Q$-conditional symmetries of the form
\be \label{2}
 {Q}= \ \partial_t+\xi(t,x,U)\partial_x+\eta(t,x,U)\partial_U,
\ee where $\xi$ and $\eta$ are unknown functions, for the RDC
equations  (\ref{1})-(\ref{1b}). We don't consider the problem of
constructing  $Q$-conditional symmetries of the form
\[
 {Q}= \partial_x+\eta(t,x,U)\partial_U,
\]
because one is equivalent (up to the known non-local transformation)
to solving the given equations (\ref{1})-(\ref{1b})
\cite{zh-lahno98}.

Now we present main results of the paper in the form of two
theorems. Note that  we search for purely
 conditional symmetry operators, which  cannot be reduced
 to Lie symmetry operators.

\bt Equation (\ref{1}) is $Q$-conditional invariant under the
operator (\ref{2})  if and only if
 it  and  the relevant operator (up to equivalent representations generated by multiplying
      on the arbitrary smooth function $M(t,x,U)$) have the  following forms:

\be\label {3}(i) \ \ \ U_t= \ [U ^{m} U_{x}]_{x}+  \lambda
U^{m}U_{x}+(\lambda_1 U^{m+1} + \lambda_2)( U^{-m}-\lambda_3),\
m\neq-1, \lambda_2 \not=0 \ee \be \label {4}\quad \quad \  Q= \
\partial_t\ +( \lambda_1 U + \lambda_2 U^{-m})\partial_U; \ee
\medskip

\be \label {5}(ii) \ \ \ U_t= \ [U ^{-1} U_{x}]_{x}+  \lambda
U^{-1}U_{x}+(\lambda_1 \ln U + \lambda_2 )(U - \lambda_3),\
\lambda_1 \not=0,\ee \be\label {6}\quad \quad \ \ {Q}=\
\partial_t\ +( \lambda_1 \ln U + \lambda_2  )U\partial_U;
\ee
\medskip
\be(iii)\label {7} \ \ \ U_t= \ [U ^{-\frac{1}{2}} U_{x}]_{x}+
\lambda U^{-\frac{1}{2}}U_{x} +\lambda_1 U+\lambda_2
U^{\frac{1}{2}}\ +\lambda_3,\ee \be \label {8}\quad \quad \quad {Q}=
\
\partial_t\ +f(t,x)\partial_x\ + 2(g(t,x)
U+h(t,x)U^\frac{1}{2})\partial_U,\ee where the function triplet ($f,
g, h$) is  the general solution of the system \be\label{9}
\ba{l}\medskip
2ff_x+f_t+fg=0,\\
\medskip f_{xx}-\lambda f_{x}-2g_{x}-fh=0,\\
\medskip
(g-\frac{\lambda_{1}}{2})(g+2f_{x})+g_{t}=0,\\
\medskip
2gh-\lambda_{1}h+2f_{x}h-\lambda_{2}f_{x}+h_{t}-\lambda g_{x}-g_{xx}=0,\\
\medskip
h^{2}-\frac{\lambda_{2}}{2}h-\lambda_{3}f_{x}+\frac{\lambda_{3}}{2}g-\lambda
h_{x}-h_{xx}=0.\ea\ee Hereafter $\lbd_1,\ \lbd_2$ and  $\lbd_3$ are
arbitrary constants. \et

It should be noted that cases (i) and (ii) with $\lambda=0$
immediately give the RD equations and the relevant symmetries
obtained in  \cite{a-h}.

 Nevertheless system (\ref{9}) contains five equations on three unknown functions
it is compatible. In fact,  the system with $f=g=0,\ \lambda_{1} =0$
is reduced to the ordinary differential equation \be \label{10}
h_{xx}+\lambda h_{x}+\frac{\lambda_{2}}{2}h-h^{2}=0 \ee  and hence
\be \label {11}{Q}= \
\partial_t\ + 2h(x)U^\frac{1}{2}\partial_U\ee
is the $Q$-conditional symmetry operator for an arbitrary non-zero
solution of (\ref{10}). Unfortunately, ODE (\ref{10}) cannot be
integrated for the arbitrary  coefficients $\lambda$ and
$\lambda_{2}$, however,    some particular  solutions can be easily
established.
   For example, setting $h=\frac{\lambda_2}{2}$, a particular
case of  $(i)$ with $m=-\frac{1}{2},\ \lambda_1=0$ is obtained.

Setting $\lambda=\lambda_{2}=0$ in (\ref{10}), we arrive at the
known ODE $h_{xx}=h^2$ with the general solution $h={\cal
W}(0,c_1,x+c_2),$ where $c_1$ and $c_2$ are arbitrary constants,
$\cal W$ is the Weierstrass function with the periods $0$ and $c_1$.
Its  simplest solution  takes the form $h=6x^{-2}$ and leads to the
known $Q$-conditional symmetry operator $Q=\partial_t\ +
12x^{-2}U^\frac{1}{2}\partial_U$ of the nonlinear diffusion equation
$U_t= \ [U ^{-\frac{1}{2}} U_{x}]_{x}$ \cite{a-h}.  However the
result derived in \cite{a-h} can be generalized as follows. One can
easily check that an arbitrary particular  solution  of (\ref{9})
with  $\lambda =0$  generates
 the  $Q$-conditional symmetry operator  (\ref{8})
 of the RD equation
 \be \label{7a}  U_t= \ [U ^{-\frac{1}{2}} U_{x}]_{x}+
\lambda_1 U+\lambda_2 U^{\frac{1}{2}}\ +\lambda_3.\ee Obviously, the
operator presented in Table 3 of   \cite{a-h} is obtainable   from
(\ref{8}) and (\ref{9}) by setting $\lambda =\lambda_1= 0$ and
$f=g=0$ but not wise versa.


\bt Equation (\ref{1b})
is $Q$-conditional invariant under the operator (\ref{2})  if and
only if
 it  and  the relevant operator (up to equivalent representations generated by multiplying
      on the arbitrary smooth function $M(t,x,U)$) have the  following forms:
\be\label {50a} (i) \ \ \ U_t= \ [U ^m U_x]_x+  \lambda
U^{m+1}U_{x}+\lambda_1 U + \lambda_2 U^{-m},\ m\neq-1,  \ee \be
\label {51}\quad \quad \ {Q}= \
\partial_t-\lbd U^{m+1}\partial_x +( \lambda_1 U + \lambda_2 U^{-m})\partial_U; \ee
\medskip
\be \label {52} (ii) \ \ \ U_t= \ [U ^{-{1\over{2}}} U_x]_x+ \lambda
U^{1\over{2}}U_{x}+(\lbd_1 U^{3\over2}+ \lambda_2 U^{1\over
2}+\lbd_3)\lf({\lbd_1\over {2 \lbd^2}} + U^{1\over2}\rg),\ee
\be\label {53}\quad \quad \ \  {Q}=\
\partial_t+\lf(-\lbd U^{1\over2}+{3\lbd_1\over2\lbd}\rg)\partial_x+(\lbd_1 U^{3\over2}+ \lambda_2 U^{1\over
2}+\lbd_3)\partial_U; \ee
\medskip
\be \label {6-ad} (iii)\ \ \ U_t =  U_{xx} + \lambda UU_x,
 \ee
\be \label{7-ad}\quad \quad \quad  Q =\partial_t +\lf({\lambda \over
2}U+q\rg)\partial_x + \lf(a+bU-{\lambda q \over 2}U^2- {\lambda^2
\over 4}U^3 \rg)\partial_U, \ee where triplet of the functions
$(a,b,q)$ is the general solution of the system
 \be \label{8-ad}
\ba{l}
\medskip
 a_t =  a_{xx} -2aq_x,\\
\medskip
b_t =  b_{xx} -2bq_x+\lambda a_x,\\
q_t =  q_{xx} -2 qq_x-2b_x ; \ea \ee
\medskip
 \be \label{9-ad} (iv)\ \ \ U_t = U_{xx} + \lambda
UU_x+\lambda_0
+\lambda_2  U^2, \quad \lambda_2\not=0,
 \ee
\be \label{10-ad}\quad \quad \quad  Q =\partial_t +\lf(-\lambda
U+{\lambda_2 \over \lambda}\rg)\partial_x + (\lambda_0
+\lambda_2U^2 )\partial_U; \ee
\medskip
\be \label{11-ad} (v)\ \ \ U_t =  U_{xx} + \lambda
UU_x+\lambda_0+\lambda_1U
+\lambda_3  U^3, \quad \lambda_3\not=0;
 \ee
\be \label{12-ad}\quad \quad \ \  Q_{i} =\partial_t +
 p_{i}U\partial_x + {3p_{i} \over
2p_{i}-\lambda} (\lambda_0+\lambda_1U
+\lambda_3 U^3 )\partial_U,\ i=1,2
\ee where $p_{i}$ are the roots of the quadratic equation
$2p^2+\lambda p+9\lambda_3- \lambda^2=0$,
  \be \label{13-ad}
      Q =\partial_t +b\partial_x + (\gamma b_{xx}-b_x U)\partial_U,
      \ee where the function $b(t, x)$ is the general  solution of the
      overdetermined system
\be \label{14-ad} \ba{l}
\medskip
  (\lambda\gamma-3)  b_{xx} +2 bb_x +b_t =0, \\
\medskip
   b_{xxx} -bb_{xx}+ \lambda_1b_x=0,\\
    \lambda\gamma b_{xxx} +b_{x}^2+ 3\lambda_1b_x+\frac{3\lambda_0}{\gamma}b =c_0, \ea \ee
    where $\gamma=\frac{\lambda}{3\lambda_3}$,$\lambda_0\lambda_1 \lambda_3\not=0$
     and $\lbd_0,\ c_0 \in \mathbb{R}.$
\et

\emph{\textbf{Remark 1.}}
The case of the Burgers equation (\ref{6-ad}) was completely
investigated  in paper \cite{ar-b-h}(see also
 a locally equivalent equation to  the Burgers equation found in \cite{ch-se-98}
and particular solutions of (\ref{8-ad}) presented  in
\cite{ames,ch-nataly-96}), while the cases $(iv)-(v)$ were obtained
in \cite{che-2006}.
\medskip

\emph{\textbf{Remark 2.}} The RDC equations\[U_t=U_{xx}+\lbd U
U_x+\lbd_0+\lbd_1 U+\lbd_2U^2\]and\[U_t=U_{xx}+\lbd U
U_x+\lbd_0+\lbd_1 U+\lbd_2U^2+\lbd_3U^3\] possess  Q-conditional
symmetry, however they are reduced to (\ref{9-ad}) and
(\ref{11-ad}), respectively, by the simple local substitutions
\cite{che-2006}.\medskip

Inserting $\lambda=0$ into (\ref{11-ad})--(\ref{12-ad}) one
immediately arrives at the well-known RD equation and
$Q$-conditional symmetry operator with cubic nonlinearities
constructed earlier \cite{se-90}-\cite{cla}.

We have also proved  that  the first two equations of (\ref{14-ad})
with  $\lambda=0$ and $b_t=0$ is equivalent to
 the equation $6b_x-2b^2+9\lambda_1=0$ derived in \cite{cla} (see P.264)
to construct  the  $Q$-conditional symmetry operator  (\ref{13-ad})
in the explicit form for  the RD  equation  with cubic nonlinearity.

Consider some equations which arise as particular cases of those
from Theorem 1 and 2 and are known in application. Equation
(\ref{3}) with $m=1$ contains as a subcase the equation \be
\label{3*}U_t= \ [U  U_{x}]_{x}+  \lambda UU_{x}+U(1- U), \ee which
is a natural generalization of the equation \be \label{3**}U_t= \
U_{xx}+  \lambda UU_{x}+U(1- U), \ee extensively studied  by  Murray
\cite [Section 11.4]{mur}. On the other hand, (\ref{3*}) is nothing
else but the porous-Fisher equation with the Burgers  convective
term
 $UU_{x}$. Note that the Murray equation (\ref{3**}) arises in Theorem 2 (see case $(iv)$ and Remark 2).
 Its $Q$-conditional symmetry was established earlier in \cite{ch-se-98} and several exact solutions
 were recently found in  \cite{che-2006}.
 Equation (\ref{7}) with $\lambda_2=-\lambda_1=-2, \lambda_3=0, t \to 2t$ takes the form
 \be\label{7*} U_t= \ (U^{\frac{1}{2}})_{xx}+  \lambda (U^{\frac{1}{2}})_{x}+ U^{\frac{1}{2}}(1- U^{\frac{1}{2}}). \ee
This equation  may be called the Murray equation with the fast
diffusion. Another analog of the Murray equation with the fast
diffusion is \be\label{7**} U_t= \ (U^{\frac{1}{2}})_{xx}+  \lambda
U^{\frac{1}{2}}U_{x}+ U^{\frac{1}{2}}(1- U^{\frac{1}{2}}), \ee which
is a particular case of equation (\ref{50a}).

  Consider the generalized Fitzhugh-Nagumo (FN) equation
  \be \label{21a-ad} U_t = U_{xx} + \lambda UU_x+\lambda_3 U(U-\delta)(1-U),\quad 0<\delta<1.
 \ee
 We remind the reader that (\ref{21a-ad}) with
  $\lambda=0$ is the famous FN equation \cite{fitzhugh} describing nerve impulse propagation. It
can be also considered as a simplification of Hodgkin-Huxley model
(see, e.g., \cite{mur2}) describing the ionic current flows for
axonal membranes. Equation
 (\ref{21a-ad}) with
  $\lambda=0$ and $\delta=1$ is the Kolmogorov-Petrovskii-Piskunov  equation,
  which firstly was investigated in  \cite{kpp-1937} (see English translation in \cite{kpp-1982})
   to describe the population
  dynamics under some restrictions on characteristic individuals.
 Equation (\ref{21a-ad}) can be reduced to the form
   \be \label{21-ad} W_t =  W_{yy} + \lambda
WW_y+\lambda_0+\lambda_1W -\lambda_3  W^3,\ee where
 \be \label{21c} \lambda_1=\lambda_3(\frac{1}{3}(\delta+1)^2-\delta), \quad \lambda_0=\lambda_3\frac{1}{3}(\delta+1)
 (\frac{2}{9}(\delta+1)^2-\delta),\ee
by the local substitution
 \be \label{21d} W(t,y)=U-\frac{1}{3}(\delta+1), \quad y=x+ \frac{\lambda}{3}(\delta+1)t. \ee
 Now one note that  equation (\ref{21-ad}) is nothing else but equation  (\ref{12-ad}) with the  new notation.
 On the other hand,  (\ref{52}) contains as a particular case the equation
 \be\label{52*} U_t= \ (U^{\frac{1}{2}})_{xx}+  \lambda U^{\frac{1}{2}}U_{x}+
 U^{\frac{1}{2}}(U^{\frac{1}{2}}-\delta)(1- U),\ \delta=\frac{9}{\lambda^2}, \ee
 which may be treated as a generalized FN equation with the fast
 diffusion. In the case $\lambda=\pm 3$, this equation may be called
 the generalized Kolmogorov-Petrovskii-Piskunov  equation
with the fast diffusion.

\newpage

\begin{center}
\textbf{3. Proofs of the theorems.}
\end{center}

\textbf{Proof of Theorem 1.} The proof of Theorem 1 and 2 is based
on the known algorithm for finding $Q$-conditional symmetry
operators (see, e.g., \cite{ch-se-98}, \cite{Fush93}). Firstly, we
apply  the local substitution

 \be\label {12} V= \cases{
\medskip
 {U^{m+1},\ m\neq-1,}
\cr
 {\ln U,\ m=-1.}}
\ee

In the cases   $\ m\neq-1$ and $m=-1$,   substitution  (\ref{12})
reduces equation  (\ref{1}) to the forms
 \be\label {13} V_{xx}=V^{n}V_{t}-\lambda V_{x}+F(V),
 \ee
(here  $n=-\frac{m}{m+1}\neq 0,\ F(V)=-(m+1)C(V^{\frac{1}{m+1}}), \
\lambda\neq0$) and \be \label {14} V_{xx}=\exp(V)V_{t}-\lambda
V_{x}+F(V), \quad F(V)=C(\exp V), \ee respectively.

The determining equations for the general RDC equation
\[V_{xx}=F_{0} (V)V_{t}+F_{1} (V)V_{x}+F_{2} (V),\]
being  $F_{i}(V), i=1,2,3$ arbitrary functions,
 have been obtained in \cite{ch-se-98} (see P.535).
In the case
 $F_0(V)=~V^{n},\
F_1(V)=-\lambda$ and $ F_2(V)=F(V)$ those equations take the form

\be\label {16}\ba{l}
\medskip
\xi_{VV}=0,\\
\medskip
\eta_{VV}=2\xi_{V}(-\lambda-\xi V^{n})+2\xi_{xV},\\
\medskip
(2\xi_{V}\eta-2\xi\xi_{x}-\xi_{t})V^{n}-\xi\eta
nV^{n-1}-\lambda\xi_{x}+3\xi_{V}F-2\eta_{xV}+\xi_{xx}=0,\\
\medskip
\eta F_{V}+(2\xi_{x}-\eta_{V})F+n\eta^{2}V^{n-1}+2\xi_{x}\eta
V^{n}+\eta_{t}V^{n}-\lambda\eta_{x}-\eta_{xx}=0.\ea \ee

Solving the first equation of (\ref{16}), we arrive at the function
$\xi=a(t,x)V+f(t,x)$ being $a(t,x)$ and $f(t,x)$ arbitrary smooth
functions at the moment.

It turns out that system  (\ref{16}) doesn't possess any
$Q$-conditional symmetry if    $a(t,x)\neq0$. So we must assume
\be\label {17} \xi=f(t,x).\ee Solving the second equation of
(\ref{16}) under condition   (\ref{17}), we arrive at
 \be \label {18}\eta=g(t,x)V+h(t,x).\ee
Taking into account
 (\ref{17}) and  (\ref{18}) the third equation of (\ref{16}) reduces to the form
 \be\label {19} (2ff_{x}+f_{t}+n f
g)V^{n}+nfhV^{n-1}-f_{xx}+\lambda f_{x}+2g_{x}=0. \ee This equation
can be splitted  with respect to the powers of $V$. One needs to
consider two cases depending on
 $n$: \medskip

(a) if  $n\neq1$ then \be\label{42}\ba{l}
\medskip 2ff_{x}+f_{t}+n fg=0,\\
\medskip fh=0,\\
\medskip f_{xx}-\lambda f_{x}-2g_{x}=0.\ea\ee

(b) if  $n=1$ then
\[\ba{l}
\medskip 2ff_{x}+f_{t}+fg=0,\\
\medskip f_{xx}-\lambda f_{x}-2g_{x}-fh=0.\ea\]

Let us consider case (a). Substituting  (\ref{17}) and  (\ref{18})
into the fourth equation of
 (\ref{16}) one arrives at  \be\label {20}\ba{c}
(gV+h)F_{V}+(2f_{x}-g)F=-nV^{n-1}(gV+h)^{2}+h_{xx} +\lambda
h_{x}+\\+(g_{xx}+\lambda
g_{x})V-(g_{t}+2f_{x}g)V^{n+1}-(h_{t}+2f_{x}h)V^{n}.\ea \ee To solve
(\ref{20}) and (\ref{42}) one needs to consider two subcases, which
follow from the second equation of  (\ref{42}), i.e. either $f=0$
or  $h=0$.

The case  $f=0$ leads to the system \be\label{45}\ba{l}\medskip f=0,\\
\medskip
g_x=0,\\
\medskip (gV+h)F_V-gF=-nV^{n-1}(gV+h)^{2}+h_{xx} +\lambda
h_{x}-g_t V^{n+1}-h_tV^n.\ea\ee

 Setting  $g~=~const,\
h=const$, we arrive at the system
\[f=0,\ g=\lambda_1^*,\ h=\lambda_2^*, \
F=(\lambda_1^{*} V+\lambda_2^{*})(\lambda_{3}-V^n),\] therefore
\be\label{43} V_{xx}=V^{n}V_{t}-\lambda V_{x}+(\lambda_1^{*}
V+\lambda_2^{*})(\lambda_{3}-V^{n}),\ee \be\label {44}
Q=\partial_{t}+(\lambda_1^{*} V+\lambda_2^{*})\partial_V.\ee
Applying substitution (\ref{12}) to the equation (\ref{43}) and
operator  (\ref{44}) we obtain case   $(i)$ of the theorem  (note
one should use new notations $\lambda_i=\frac{\lambda_i^*}{m+1},\
i=1,2$).

Now we assume that  $g\ne constant$, so that the third equation of
(\ref{45}) can be reduced to the form
\be\label{50}\lf(V+\frac{h}{g}\rg)F_V-F=-ngV^{n-1}\lf(V+\frac{h}{g}\rg)^{2}+\frac{h_{xx}
+\lambda h_{x}}{g}-\frac{g_t}{g} V^{n+1}-\frac{h_t}{g}V^n.\ee

It turns out that the last equation can be satisfied only under
condition  $\frac{h}{g}=const$ (see the proof below). Setting
$\frac{h}{g}=constant$ into (\ref{50}) and making the relevant
calculations,  we obtain only Lie symmetry operators and a
particular case of  operator (\ref{44}) and  equation (\ref{43}).
For example, if  ${h\over g}=0$ then  the system
 \be\label{46}\ba{l}
\medskip h=0,\\
\medskip 2ff_{x}+f_{t}+n fg=0,\\
\medskip f_{xx}-\lambda
f_{x}-2g_{x}=0,\\ \medskip gVF_{V}+(2f_{x}-g)F=-nV^{n+1}g^{2}
+(g_{xx}+\lambda g_{x})V-(g_{t}+2f_{x}g)V^{n+1}\ea\ee
 is obtained, which can be easily  solved and its general solution has the form
  \[f=\frac{c_1\exp(\lambda_1nt)}{c_2\exp(\lambda_1nt)+1},\
g=-\frac{\lambda_1}{c_2\exp(\lambda_1nt)+1},\ h=0,\ F=\lambda_1
V^{n+1}+\lambda_2 V,\] where $c_k \in \mathbb{R},\ k=1,2.$ Hence we
arrive at the RDC equation
\[V_{xx}=V^nV_t-\lambda V_x+\lambda_1V^{n+1}+\lambda_2 V\]
and  the operator
\be\label{50*}Q=\partial_t+\frac{c_1\exp(\lambda_1nt)}{c_2\exp(\lambda_1nt)+1}
\partial_x-\frac{\lambda_1 V}{c_2\exp(\lambda_1nt)+1}\partial_V.\ee
However one can establish by multiplying (\ref{50*}) on the function
$M(t,x,U)=1+c_2\exp(\lbd_1 n t)$ that the last operator is nothing
else but the Lie symmetry operator (see case  8 of Table 1 in
\cite{ch-se-98}).

Let us prove that  $\frac{h}{g}=const$. By differentiating  equation
(\ref{50}) with respect to the variables $x$ and  $t$
one obtains two equations. Assuming
$(\frac{h}{g})_t(\frac{h}{g})_x=0$, one easily arrives at the
condition
$\frac{h}{g}=const$.

Consider the case  $(\frac{h}{g})_t(\frac{h}{g})_x\neq0$. By
differentiating  equation (\ref{50}) with respect to the variables
$x$ we arrive at the equation

\[F_V=-\frac{1}{h_x}(2ngh+h_t)_x
V^n-2nhV^{n-1}+\frac{h_{xxx}+\lambda h_{xx}}{h_x}.\] Since the
function $V^n,\ V^{n-1}$ and  $1$ on the right-hand-site  are
functionally independent (we consider the case $n\ne1$) their
coefficients must by constants. It means that
 $2nh=const$ so that  $h_x=0$. Taking now two last equations of   (\ref{42}),
 one easily establish that
$g_x=0$, i.e. we arrive at the contradiction:  $(\frac{h}{g})_x =0$.

Consider the case  (b). Substituting  (\ref{17}) and  (\ref{18})
into the fourth equation of  (\ref{16}), we arrive at  (\ref{20})
with $n=1$. Dealing with this equation in the same way as above (see
the case (a)) we obtain  equation \be\label {21}
V_{xx}=VV_{t}-\lambda
V_{x}+\lambda_{1}^{*}V^{2}+\lambda_{2}^{*}V+\lambda_{3}^{*}, \ee and
the operator \be\label {22}
Q=\partial_{t}+f(t,x)\partial_{x}+(g(t,x)V+h(t,x))\partial_{V},\ee
were the triplet  ($f,\ g,\ h$) are the general solution of
(\ref{9}). Applying formula  (\ref{12}) with $m\not=-1$ we obtain
the case  $(iii)$ of the theorem (note one should use new notations
$\lambda_{i}=-2\lambda_{i}^{*},\ i=1,2,3$).

Finally, we analyze equation (\ref{14}), which is locally equivalent
to the RDC (\ref{1}) with $m=-1$. Using again the determining
equations, which  have been obtained in \cite{ch-se-98}  to find
operators of the $Q$-conditional symmetries  (\ref{4}), we arrive at
the following system \be\label {23} \ba{l}
\medskip \xi_{VV}=0,\\
\medskip
\eta_{VV}=2\xi_{V}(-\lambda-\xi \exp V)+2\xi_{xV}, \\
\medskip
(\xi_{t}+2\xi\xi_{x}-2\xi_V\eta+\xi\eta)\exp V+\lbd\xi_x-3\xi_{V}F+2\eta_{xV}-\xi_{xx}=0,\\
\medskip
\eta F_{V}+(2\xi_{x}-\eta_{V})F+(\eta^{2}+2\xi_{x}\eta
+\eta_{t})\exp V-\lambda\eta_{x}-\eta_{xx}=0,\ea \ee where $\xi$,
$\eta$ and $F$ are yet-to-be determined functions. Solving the first
and second equations of this system we establish that the functions
$\xi$ and $\eta$ must be given by formulas  (\ref{17}) and
(\ref{18}), respectively, otherwise Q-conditional symmetry doesn't
exist. Substituting (\ref{17}) and (\ref{18}) into the third
equation of (\ref{23}) we obtain the equation
 \[(fh
+f_{t}+2ff_{x})\exp V+(fg)V\exp V+\lambda f_{x}+2g_{x}-f_{xx}=0.\]
Since the functions  $f,\, g$ and $ h$ don't depend on  $V$, one can
split this equation with respect to $\exp V$ and  $V\exp V$ and
obtain the system \be\label {24} \ba{l}
\medskip f h +f_{t}+2ff_{x}=0,\\
\medskip
fg=0, \\
\medskip \lambda f_{x}+2g_{x}-f_{xx}=0.\ea
\ee

Substituting  (\ref{17}) and  (\ref{18}) into the fourth equation of
(\ref{23}) we arrive at the equation
 \be\label {25} \ba{cc}
(gV+h)F_{V}+(2f_{x}-g)F=-(gV+h)^2\exp V+h_{xx} +\lambda
h_{x}+\\+(g_{xx}+\lambda g_{x})V-(g_{t}+2f_{x}g)V\exp
V-(h_{t}+2f_{x}h)\exp V.\ea \ee Now we apply to
 (\ref{25}) the same approach, which has been used for solving equation  (\ref{20}).
 Thus taking into account system  (\ref{24}), we  obtain finally the expressions
\[F=(\lambda_{1} V+\lambda_{2})(\lambda_{3}-\exp V),\ f=0,\ g=\lambda_{1},\ h=\lambda_{2},\]

which lead to the equation
 \be\label {26} V_{xx}=\exp(V)V_{t}-\lambda V_{x}+(\lambda_{1}
V+\lambda_{2})(\lambda_{3}-\exp V)\ee  and the operator
 \be\label {27}
Q=\partial_{t}+(\lambda_{1} V+\lambda_{2})\partial_{V}.\ee

Applying substitution   (\ref{12}) with $m=-1$ to  (\ref{26}) and
(\ref{27}) one obtains the case  $(ii)$ of the Theorem 1.

The proof  is now completed.

\textbf{Proof of Theorem 2.}
 First of all we note that all $Q$-conditional symmetries of equation  (\ref{1b}) with $m=0$
 were found in the recent paper \cite{che-2006} so that  the restriction  $\ m\neq 0$ is assumed below.
 We again use substitution
(\ref{12}), which reduces equation  (\ref{1b}) to the form
 \be \label{60}
V_{xx}=V^nV_{t}-\lambda V^{n+1} V_x+F(V),\ee (here
$n=-\frac{m}{m+1}\ne0,-1,\ F(V)=-(m+1)C(V^{\frac{1}{m+1}})$) if  $\
m\neq-1$ \ and to the form
 \be \label {61} V_{xx}=\exp(V)V_{t}-\lambda \exp(V)
V_{x}+F(V), \quad F(V)=C(\exp V)\ee if  $m=-1$.

Consider equation (\ref{60}). Using the general form of the
determining equations obtained in \cite{ch-se-98} one easily arrives
at the following system \be\label {62} \ba{l}
\medskip \xi_{VV}=0,\\
\medskip
\eta_{VV}=2\xi_{V}(-\lambda V^{n+1}-\xi V^{n})+2\xi_{xV},\\
\medskip
\eta F_{V}+(2\xi_{x}-\eta_{V})F+n\eta^{2}V^{n-1}+2\xi_{x}\eta
V^{n}+\eta_{t}V^{n}-\lambda V^{n+1}\eta_{x}-\eta_{xx}=0,\\
\lbd \xi_x V^{n+1}+\biggr((-2\xi_V+\lbd
(n+1))\eta+2\xi\xi_{x}+\xi_{t}\biggr)V^{n}+\xi\eta
nV^{n-1}-3\xi_{V}F+\\+2\eta_{xV}-\xi_{xx}=0.\ea \ee to find the
function $\xi,\ \eta$ and  $F$. In the case of equation  (\ref{61})
that system takes the form \be\label {63} \ba{l}
\medskip \xi_{VV}=0,\\
\medskip
\eta_{VV}=-2\xi_{V}(\lambda +\xi) \exp V+2\xi_{xV}, \\
\medskip
\biggr(\xi_t+2\xi\xi_{x}+(\lbd+\xi-2\xi_V)\eta+\lbd\xi_x\biggr)\exp V-3\xi_{V}F+2\eta_{xV}-\xi_{xx}=0,\\
\medskip
\eta F_{V}+(2\xi_{x}-\eta_V)F+(\eta^2+2\xi_{x}\eta
+\eta_{t}-\lambda\eta_x)\exp V-\eta_{xx}=0.\ea \ee

Taking into account the first equation in (\ref{62}) and (\ref{63}),
one establish that there are only three possibilities for the
functions $\xi$ and $ \eta$ :

\[(a)\ \xi= \lbd_1^*V+\lbd_2^*,\  \, \eta=\eta(V), \quad \lbd_1^*, \lbd_2^*\in \mathbb{R}, \]
\be\label{133}(b)\ \xi=f(t,x),\ \eta=g(t,x)V+h(t,x),\ee
\[(c)\ \xi=a(t,x)V+f(t,x),\  \eta=\eta(t,x,V), \quad a(t,x)\ne0.\]
In the case $(c)$ function $\eta(t,x,V)$ takes the forms \be\label
{133a} \eta= \cases{
\medskip-\frac{2a(a+\lbd)}{(n+2)(n+3)}V^{n+3}-\frac{2af}{(n+1)(n+2)}V^{n+2}+a_x
V^2+g(t,x)V+h(t,x),\  n\ne-2,-3, \cr\medskip
-2a(a+\lbd)V\ln V+2af\ln V+a_xV^2+(2a(a+\lbd)+g(t,x))V+h(t,x),\
n=-2, \cr
2a(a+\lbd)\ln V-afV^{-1}+a_xV^2+g(t,x)V+h(t,x),\  n=-3; }\ee for the
system (\ref{62})  and
\[\eta=-2a^2V\exp V-2a(\lbd+f-2a)\exp V+a_xV^2+g(t,x)V+h(t,x) \]
for the system (\ref{63}).

Consider  case (a) and system  (\ref{62}). Since the function $\xi$
is the linear function, the general solution of the second equation
of (\ref{62})  is the function \be\label{65}
\eta=-\frac{2\lbd_1^*(\lbd+\lbd_1^*)}{(n+2)(n+3)}V^{n+3}-\frac{2\lbd_1^*\lbd_2^*}{(n+1)(n+2)}V^{n+2}+\lbd_3^*V+\lbd_4^*,\ee
where  $\lbd_3^*, \lbd_4^*\in \mathbb{R},\ n\ne-2,-3.$ Substituting
(\ref{65}) into the fourth equation of  (\ref{62}), one obtains
\be\label{66}
F=\frac{\eta}{3\lbd_1^*}\biggr[\biggr(\lbd_1^*(n-2)+\lbd(n+1)\biggr)V^n+n\lbd_2^*V^{n-1}\biggl],\
\ee  if $\lbd_1^*\ne0$ (the case $\lbd_1^*=0$ leads only to the Lie
symmetry operators). Substituting  (\ref{66}) into the third
equation of system (\ref{62}) one arrives at the expression
\be\label{67} \eta\lf[\biggr(1+{1\over
3\lbd_1^*}(\lbd_1^*(n-2)+\lbd(n+1)\biggr)V^{n-1}+(n-1)\lbd_2^*V^{n-2}\rg]=0.\ee


The first possibility is
 \be\label{67a}\lf(1+{1\over
3\lbd_1^*}(\lbd_1^*(n-2)+\lbd(n+1)\rg)V^{n-1}+(n-1)\lbd_2^*V^{n-2}=0\ee
while the second one is $\eta =0$.

Splitting (\ref{67a}) with respect to the different powers of $V$
one obtains the system \be\label{70}\ba{l}(\lbd_1^*+\lbd)(n+1)=0,\medskip \\
\medskip (n-1)\lbd_2^*=0.\ea \ee The first equation immediately
gives  $\lbd_1^*=-\lbd$ because $n\ne-1$. The second equation  of
(\ref{70}) can be transformed into an equality if $\lbd_2^*=0$ or
$n=1$. Hence  we obtain the equation \be\label{73}V_{xx}=V^nV_t-\lbd
V^{n+1}V_x-(\lbd_3^*V+\lbd_4^*)V^n,\ee and the relevant
$Q$-conditional symmetry operator \be\label{74}Q=\partial_t-\lbd
V\partial_x+(\lbd_3^*V+\lbd_4^*)\partial_V \ee if  $\lbd_2^*=0$, and
the equation
 \be\label{71} V_{xx}=VV_t-\lbd
V^2V_x-\frac{\lbd_2^*+3\lbd V}{3\lbd}\lf({1\over3}\lbd_2^*\lbd
V^3+\lbd_3^*V+\lbd_4^*\rg)\ee and the relevant operator
 \be\label{72} Q=\partial_t+(-\lbd
V+\lbd_2^*)\partial_x+\lf({1\over3}\lbd_2^*\lbd
V^3+\lbd_3^*V+\lbd_4^*\rg)\partial_V,\ee if $n=1$. Finally, applying
substitution   (\ref{12}) with $m \not =-1$ to  (\ref{73}) --
(\ref{72}) and introducing new notations $\lbd_3^*=\lbd_1(m+1),\
\lbd_4^*=\lbd_2(m+1)$ in the case of  equation (\ref{73}) and
$\lbd_2^*=\frac{3\lbd_1}{2\lbd},\ \lbd_3^*={\lbd_2\over2},\
\lbd_4^*={\lbd_3\over2}$  in the case  equation (\ref{71}), we
arrive exactly at the items  $(i)$ and  $(ii)$
 of the Theorem 2.

It turns out that the special values   $n=-2 $ and $ n=-3$ (see
(\ref{133a})) lead only to particular cases of equation  (\ref{73})
and operator  (\ref{74}). Moreover the case $\eta=0$ again leads to
the equation   (\ref{73}) and operator (\ref{74}) with
$\lbd_3^*=\lbd_4^*=0$.


To complete the investigation of case (a) one needs to consider
system (\ref{63}). Integrating the second equation one easily
obtains \be\label{75}
\eta=2\lbd_1^*(-\lbd_1^*V+2\lbd_1^*-\lbd-\lbd_2^*)\exp
V+\lbd_3^*V+\lbd_4^*,\ee where  $\lbd_1^*, \lbd_2^*, \lbd_3^*,
\lbd_4^*\in\mathbb{R}.$ Inserting $\xi= \lbd_1^*V+\lbd_2^*$ and
(\ref{75}) with $\lbd_1^*\ne0$ into the third equation of
(\ref{63}), we find \be\label{76} F={\eta \exp
V\over3\lbd_1^*}\lf(\lbd_1^*V+\lbd+\lbd_2^*-2\lbd_1^*\rg),\
\lbd_1^*\ne0.\ee Substituting (\ref{76}) into the fourth equation of
(\ref{63}), we arrive at the condition \be\label{77}
\eta\lf(1+{1\over3\lbd_1^*}(\lbd_1^*V+\lbd+\lbd_2^*-\lbd_1^*)\rg)\exp
V=0,\ee which can be  satisfied  only in the case $\eta=0$. However
it leads to the requirement $\lbd_1^*=0$ (see (\ref{75})) what
contradicts to (\ref{76}).

Finally,   system  (\ref{63}) with $\lbd_1^*=0$ takes the form
\be\label{78}\ba{l}\xi=\lbd_2^*,\medskip\\ \eta=\lbd_3^*V+\lbd_4^*,\medskip\\(\lbd_3^*V+\lbd_4^*)(\lbd_2^*+\lbd)=0,\medskip\\
(\lbd_3^*V+\lbd_4^*)F_V-\lbd_3^*F=-(\lbd_3^*V+\lbd_4^*)^2\exp
V.\ea\ee Solving  (\ref{78}) with respect to the function $F$, one
obtains the equation \be\label{79}V_{xx}=\exp(V)V_t-\lbd
\exp(V)V_x+(\lbd_3^*V+\lbd_4^*)(\lbd_5^*-\exp V)\ee and the relevant
$Q$-conditional symmetry operator \be\label{80} Q=\partial_t-\lbd
\partial_x+(\lbd_3^*V+\lbd_4^*)\partial_V.\ee
Applying substitution   (\ref{12}) with $m  =-1$ to  (\ref{79}) and
(\ref{80}) and introducing new notations
 $\lbd_3^*=\lbd_1,\ \lbd_4^*=\lbd_2,\
\lbd_5^*=\lbd_3,$  we obtain the RDC equation \be \label{54}
U_t=[U^{-1}U_x]_x+\lbd U_x+(\lbd_1 \ln U+\lbd_2)(U-\lbd_3)\ee and
the Q-conditional operator
 \be \label{55}Q=\ \partial_t-\lbd
\partial_x+(\lbd_1 \ln U+\lbd_2)U\partial_U. \ee
However equation (\ref{54}) and operator (\ref{55}) are reduced to
the reaction-diffusion equation
\be\label{54a}U_t=[U^{-1}U_y]_y+(\lbd_1 \ln U+\lbd_2)(U-\lbd_3)\ee
and the operator \be\label{55a} Q=\ \partial_t+(\lbd_1 \ln
U+\lbd_2)U\partial_U \ee by the local substitution $y=x+\lambda t$.
Note that (\ref{54a}) and (\ref{55a}) were previously derived in
\cite{a-h}.

Consider case (b). It turns out that this case leads only to Lie
symmetry operators. Indeed the first and the second equations of
system
  (\ref{62}) are automatically satisfied. Substituting (\ref{133})
  into the fourth equation of
(\ref{62}), one obtains \be\label{80a}
\lbd\biggr((n+1)g+f_x\biggl)V^{n+1}+\biggr(\lbd(n+1)h+f_t+2ff_x+nfg\biggl)V^n+nfhV^{n-1}+2g_x-f_{xx}=0.\ee

Now one needs to consider two subcases:  $n\ne1$ and $n=1$ (we
remind also the reader that $n\ne -1, 0$). Assuming $n\ne1$ and
splitting  (\ref{80a}) with respect to the different powers of $V$,
we obtain the system
 \be\label{129}\ba{l}\medskip(n+1)g+f_x=0,\\ \medskip\lbd(n+1)h+f_t+2ff_x+nfg=0, \\ \medskip fh=0,
\\ \medskip 2g_x-f_{xx}=0.\ea\ee
The third equation of (\ref{129}) leads to $f=0$ or $h=0$. Setting
$f=0$ one immediately arrives at   $g= h=0$ since $n+1\ne0$ and
$\lbd\ne0$. Hence we obtain a particular case (a).  Setting $h=0 $
and using  the first and the fourth equations of  (\ref{129}) we
obtain  \[g=g(t),\ f=-(n+1)g(t)x+\varphi(t),\] where the functions
$g(t)$ and $\varphi(t)$ must be determined. Substituting $g$ and $f$
into the second equation of  (\ref{129}) and splitting with respect
to  $x$, the ODE system
\[ \ba{l}\medskip g_t-(n+2)g^2=0,\\ \varphi_t-(n+2)\varphi
g=0\ea\] with the general solution \be\label{130}
g={-1\over{(n+2)t+c_1}},\ \varphi={c_2\over{(n+2)t+c_1}},\
c_k\in\mathbb{R},\ k=1,2\ee is obtained. To find the function $F(V)$
we solve the third equation of (\ref{62}), taking into account the
expressions found for  $f, \, g$ and  $h$, and obtain $F=\lbd_1
V^{2n+3}$. Hence we arrive at the equation
\be\label{131}V_{xx}=V^nV_t-\lbd V^{n+1}V_x+\lbd_1 V^{2n+3}\ee and
the operator
\be\label{132}Q=\partial_t+\frac{(n+1)x+c_2}{(n+2)t+c_1}\partial_x-\frac{V}{(n+2)t+c_1}\partial_V.\ee
However the operator  (\ref{132}) is not purely $Q$-conditional one
since it is equivalent to the operator \be\label{132a}
D=\biggr((n+2)t+c_1\biggl)\partial_t+\biggr((n+1)x+c_2\biggl)\partial_x-V\partial_V.\ee
Applying substitution   (\ref{12}) with $m  \not=-1$ to  (\ref{131})
and   (\ref{132a}) one arrives at the equation and a linear
combination of Lie symmetry operators listed in the case 7 of Table
1  \cite{ch-se-98}.

 Assuming $n=1$, we obtain only a particular case of
 (\ref{131}) and  (\ref{132}).

 Solving system  (\ref{63}) we again  find only Lie symmetry operators of  equation  (\ref{61}) with
$F=~\lbd_1+\lbd_2 \exp V $  and  $F=\lbd_1 \exp(\lbd_2V)$.

Dealing in a quite similar way with the case  (c) we have
established that this case doesn't produce any new $Q$-conditional
operators.

The proof of Theorem 2 is now completed.

\begin{center}
{\textbf{4. Conclusions.}}
\end{center}

In this paper, Theorems 1 and 2 giving a complete description of
$Q$-conditional symmetries of the nonlinear RDC equations (\ref{1})
-- (\ref{1b}) are proved. It should be stressed that all
$Q$-conditional symmetry operators listed in Theorems 1--2 contains
the same  nonlinearities with respect to  the dependent variable $U$
as the relevant RDC equations.
 Analogous results were earlier obtained   for
single reaction-diffusion  equations \cite{Fush93}, \cite{se-90},
\cite{nucci1},
 \cite{cla}.

However, we note that there is the essential difference between RDC
equations (\ref{1}) -- (\ref{1b})  and the relevant RD equation
\[U_t=[U^mU_x]_x+C(U).\] For example, the Murray type  equation (\ref{9-ad})
  admits the $Q$-conditional symmetry
 (\ref{10-ad}),
while the RD equation with this term, i.e. the Fisher type equation
\[U_t = U_{xx} + \lambda_0+\lambda_1 U+\lambda_2  U^2, \quad  \lambda_2\not=0
\]
 does not possess one. Similarly, the RDC equation (\ref{52}) possessing  the $Q$-conditional symmetry
 (\ref{53}) doesn't has an analog among  reaction-diffusion  equations with the diffusivity $U^{-{1 \over 2}}$.

The RDC equations listed in Theorems 1 and 2 contain several
well-known equations arising in applications and their direct
generalizations. In the particular case, the Murray equation
(\ref{3**}), its porous  analog (\ref{3*}) and its analogs
(\ref{7*})-(\ref{7**}) with the fast diffusion; the Fitzhugh-Nagumo
equation \cite{fitzhugh}  with the convective term
  \be \label{21a} U_t = U_{xx} + \lambda UU_x+\lambda_3 U(U-\delta)(1-U),\quad 0<\delta<1,
 \ee
 and its analog (\ref{52*}) with the  fast diffusion;
 the  Kolmogorov-Petrovskii-Piskunov  equation \cite{kpp-1937} with the convective term
  \be \label{21aa} U_t = U_{xx} + \lambda UU_x+\lambda_3 U(1-U)^2,
 \ee
and  the Newell-Whitehead equation \cite{new-wh}
 with the convective term
  \be \label{21b} U_t = U_{xx} + \lambda UU_x+\lambda_3 U^3- \lambda_1 U.
 \ee

A further generalization of the RDC equations (\ref{1}) and
(\ref{1b}) reads as
 \be \label{1-ad} U_t=[U^mU_x]_x+\lambda
U^{n}U_x+C(U), \quad \lambda \not=0, \ee
 where
 $m$ and $ n$ are   arbitrary constants. It turns out that all possible $Q$-conditional symmetries
 of (\ref{1-ad}) coincide with those presented in Theorems 1 and 2. In other words,
 if a nonlinear equation of the form (\ref{1-ad}) admits a $Q$-conditional symmetry operator
 ( not Lie symmetry !) then  either $n=m$, or $n=m+1$. The work is in progress
 on the complete description of $Q$-conditional symmetry of the RDC equation
 \[ U_t=[\exp(mU)U_x]_x+\lambda
\exp(nU)U_x+C(U), \quad \lambda \not=0. \]

It is well-known that new $Q$-conditional symmetries don't guarantee
the construction of  exact solutions, which cannot be obtained by
the Lie machinery (see non-trivial examples in \cite{ch-96,ch98}).
In the second part of this work  we  will  demonstrate  that the
$Q$-conditional symmetries obtained above can be successfully
applied for constructing new non-Lie  solutions. Moreover we will
show that  some of those  solutions have remarkable properties. In
the particular case, it will be shown that those solutions  may
satisfy typical boundary conditions arising in mathematical biology.

\end{document}